\documentstyle[sprocl,psfig]{article}

\def\apj{ApJ}
\def\apjl{ApJ}
\def\aj{AJ}
\def\mnras{MNRAS}
\def\araa{ARA\&A}

\def\al{\alpha}

\def\be{\begin{equation}}
\def\ee{\end{equation}}
\def\bea{\begin{eqnarray}}
\def\eea{\end{eqnarray}}

\begin{document}


\title{CONSTRAINTS ON THE RADIAL MASS DISTRIBUTION OF DARK MATTER
HALOS FROM ROTATION CURVES}

\author{STACY S. MCGAUGH}

\address{Department of Astronomy, University of Maryland, College Park, MD
\\E-mail: ssm@astro.umd.edu} 


\maketitle\abstracts{Rotation curves place important constraints on the
radial mass distribution of dark matter halos, $\rho(r)$.  At large radii, rotation
curves tend to become asymptotically flat.  For $\rho(r) \propto r^{\al}$,
this implies $\al \approx -2$, which persists as far out as can be measured. 
At small radii, the data strongly prefer dark matter halos with constant density
cores ($\al \approx 0$) over the cuspy halos ($\al \le -1$) predicted
by cosmological simulations.  As better data have been obtained, this cusp-core
problem has become more severe.
}

\section{Introduction}

Rotation curves played an essential role in establishing the need for dark
matter halos (Rubin, Thonnard, \& Ford 1978; Bosma 1978).  Assuming
Newtonian gravity holds, dark matter is required in all spirals to explain the
asymptotically flat rotation curves observed at large radii.  The need for dark
matter is less obvious at small radii.

Rotation curves provide strong constraints on the radial shape of
the gravitational potential of galaxies.  This translates to a constraint on the
distribution the various mass components, both luminous and dark.  Though the
constraint on the potential is strong, there can be degeneracies between the
contributions of the various mass components.  In particular, the mass of the
stellar disk can often be traded off against the mass in dark matter, leaving
considerable room for differences of opinion about disk masses and halo
distributions.  

This disk-halo degeneracy has plagued the field for a good while now. 
Consequently, there is at present a wide diversity of opinion as to how
important the luminous and dark mass are at small radii.  These range between
two easily identifiable extremes: maximal disks and cuspy halos.  In between
these is a range occupied by models with stellar mass-to-light ratios which are
plausible for the composite stellar populations of spiral galaxies.  Which of
these various options one prefers is largely a matter of how one weighs the
evidence.  Dynamicists find much to recommend maximal disks.  Cosmologists
expect cuspy dark matter halos, for which the disk mass must be minimal. 
Those who study stellar populations prefer intermediate disk masses. 

Ideally, data should dictate some resolution to this difference of opinion.  Low
surface brightness (LSB) galaxies can play an important role in this.  The
properties of these objects differ from those of high surface brightness (HSB)
galaxies in the sense that the mass discrepancy is more pronounced at all radii.
This goes a long way towards breaking the long standing disk-halo degeneracy.

\section{Halo Models}

The subject of this conference is halo shapes.  Ideally, we would like a complete
map of the dark matter distribution in galaxies, $\rho(r,\theta,\phi)$.  Unlike
many of the other contributions in these proceedings, the data discussed here
provide little handle on the distribution in $\theta$ and $\phi$.   I therefore
restrict my comments to the azimuthally averaged $\rho(r)$.  I will focus on the
limits of small and large radii, where the density distribution may be
approximated as a power law: $\rho(r) \propto r^{\alpha}$.

There are two basic halo models which are widely considered: those with
constant density cores ($\alpha \approx 0$) at small radii, and those with cusps
($\al \le -1$).  The pseudo-isothermal halo, with a constant density core rolling
over to $\al = -2$ at large radii has traditionally been used in fitting rotation
curves.  Cuspy halos are motivated by cosmological simulations in which dark
matter halos are found to take a form with a steep central cusp ($\al = -1$:
Navarro, Frenk, \& White 1997 or $\al = -1.5$: Moore et al.\ 1999), rolling over
to $\al = -3$ at large radii.  These models are different in both limits, and data
can distinguish between them.

\section{Large Radii}

It is well established (e.g., Sofue \& Rubin 2001) that at large radii, rotation
curves tend to become flat.  That is, $V \rightarrow$ constant.  This implies
$\alpha \approx 2$.

This is a remarkable fact, best illustrated by some historical cases where the
rotation can be traced to very large radii by 21 cm emission.  For
example, Fig. 1 shows the case of NGC 2403 (Begeman, Broeils, \& Sanders
1991), where the rotation curve is observed to remain flat out to 10 scale
lengths.  By this point, the luminous mass is totally encompassed, and its
contribution to the rotation has fallen far below the observations. 

\begin{figure}[t]
\psfig{figure=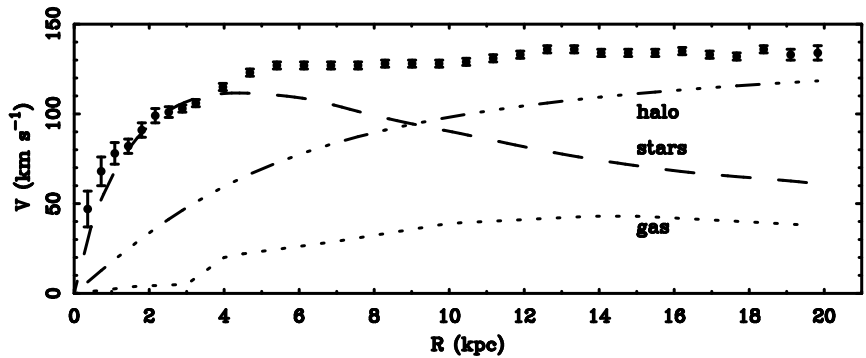,width=4.5in}
\caption{The rotation curve of NGC 2403 (points), measured out to $\sim 10$
scale lengths.  The contributions of visible stars (for $\Upsilon_*^B = 1.5\;
M_{\odot}/L_{\odot}$) and gas are shown as dashed and dotted lines,
respectively.  The rotation curve becomes approximately flat at large radii, and
stays that way indefinitely.  Note that there is little room for dark matter
at small radii, and that the halo contribution (dot-dashed line:
Begeman et al.\ 1991) is still rising at the last measured point.
\label{n2403MD}}
\rule{5cm}{0.2mm}\hfill\rule{5cm}{0.2mm}
\end{figure}

This basic observation remains poorly understood.  While a flat rotation curve is
generally presumed to be the signature of the dark matter halo, it could well be
an indication of new physics (e.g., Milgrom 1983).  Even setting this possibility
aside, models for dark matter halos do a remarkably poor job of explaining the
flatness of rotation curves which motivated them.

The trick is in understanding why rotation curves remain flat for as far as they
do.  The pseudo-isothermal model was designed to do this, but in practice the
disk contribution in HSB galaxies is large enough that the halo contribution
is often still rising to the last measured point (Fig. 1).  The observed flatness is
a fine-tuned combination of falling disk and rising halo.  This is even more true
for cuspy halos, which do not themselves produce flat rotation curves, as $\al
\ne -2$.  

There are enough parameters available to any halo model that it is usually
possible to obtain a fit to data.  However, the inverse is not true.  If one tries to
build {\it ab initio\/} disk+halo models, the resulting rotation curves have more
curvature than do real galaxies (McGaugh \& de Blok 1998).  Flat rotation curves
do not arise naturally. 

\section{Small Radii}

While the strongest constraints at large radii are most commonly obtained from
extended 21 cm measurements, those at small radii are provided by data of high
spatial resolution.  For HSB galaxies, some excellent CO data exist (Sofue \&
Rubin 2001), while for LSB galaxies (which are notoriously difficult to detect in
CO) the best constraints are provided by H$\al$ data (Swaters, Madore, \&
Trewhella 2000; McGaugh, Rubin, \& de Blok 2001; de Blok \& Bosma 2002). 
These latter have seeing limited ($\sim 1"$) resolution, an order of magnitude
improvement over early studies of these objects in HI (van der Hulst et al.\
1993; de Blok, McGaugh, \& van der Hulst 1996).

High resolution observations of the shapes of rotation curves are useful for
mapping out the shape of the potential at small radii.  There are two issues to
which the initial rate of rise of the rotation curve is particularly important: 
maximal disks and cuspy halos.  I will discuss first the issue of disk mass, then
the test for halo models (core or cusp). 

\subsection{Disk Masses}

The mass of the stars in the disk is of interest in itself, and must be
constrained in order to estimate the remaining dark mass.  The luminosity and
distribution of the stars themselves are well observed; the parameter of merit
is the stellar mass-to-light ratio $\Upsilon_*$.  This can vary over a wide
range, from the pathological limit of zero (minimum disk) up to a maximum
allowed by the data (maximum disk).  

In general, maximum disk works well to explain the shape of the inner rotation
curve, especially in HSB galaxies (e.g., Palunas \& Williams 2000).  A natural
inference is that the disk does in fact dominate where its rotation curve
matches the observed one well.  There is considerable ancillary evidence to
support the supposition that disk masses are nearly maximal for HSB galaxies
(e.g., Sellwood 1998), including our own Milky Way (Gerhard 2000).  Many
aspects of the observed dynamics appear to require that the disk contain
significant mass, a conservative lower limit being half of the total mass at two
scale lengths.

\begin{figure}[t]
\psfig{figure=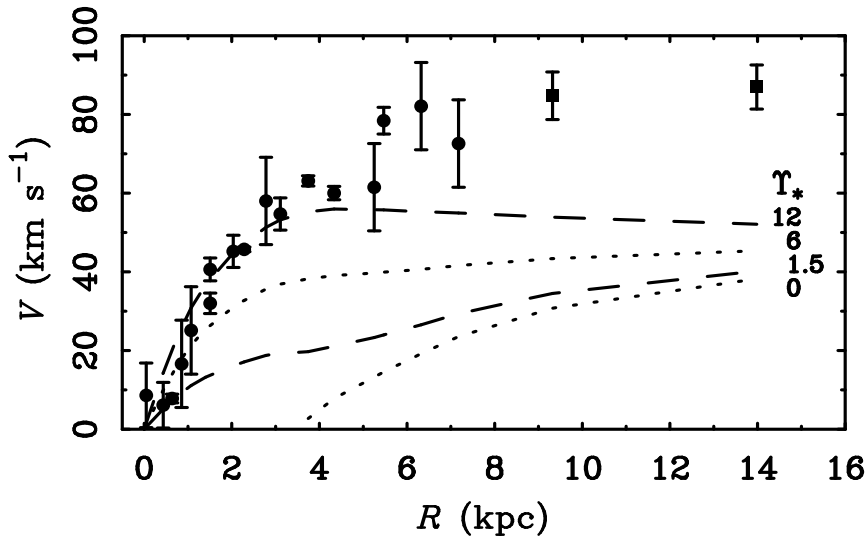,width=4.5in}
\caption{The rotation curve of the LSB galaxy
F583-1.  Circles are the H$\al$ data of McGaugh, Rubin, \& de Blok (2001);
squares are the HI data from de Blok, McGaugh, \& van der Hulst (1996).  Lines
show the baryonic disk contribution for various assumed mass-to-light ratios. 
These range from zero (no stellar mass: the lowest line is gas only) up to
$\Upsilon_*^R = 12$ (topmost line).  For reasonable stellar population 
mass-to-light ratios ($\Upsilon_*^R \approx 1.5$), LSB galaxies are halo
dominated down to small radii ($< 1$ kpc in this case).  A substantially larger
(maximal) disk mass is consistent with the data, but requires an absurdly large
mass-to-light ratio.  This just shifts some of the dark matter from halo to disk,
and a dominant halo is still required.
\label{F5831}}
\rule{5cm}{0.2mm}\hfill\rule{5cm}{0.2mm}
\end{figure}

Though maximum disk works well in HSB galaxies, it makes less sense in LSB
galaxies.  Figure 2 illustrates several possible disk masses for the LSB galaxy
F583-1, ranging from $\Upsilon_* = 0$ up to 12 in the $R$-band.  A reasonable
stellar population value of $\Upsilon_*^R = 1.5\;M_{\odot}/L_{\odot}$ makes a
small contribution to the total rotation everywhere.  If such a stellar population
model is anywhere close to the right number, as it appears it must be for
consistency with the baryonic Tully-Fisher relation (McGaugh et al.\ 2000; Bell
\& de Jong 2001), then the halo dominates this galaxy down to well within 1
kpc. 

Considering the dynamical evidence alone, one can certainly contemplate a much
higher disk mass.  The maximum disk $\Upsilon_* = 6.5$ in this case (de Blok,
McGaugh, \& Rubin 2001) if one does not allow the disk contribution to exceed
the smooth envelope of the data.  If one allows a little bit of overshoot and
tries to `fit' as much of the data with the disk as possible (e.g., Palunas \&
Williams 2000), then $\Upsilon_* \approx 12$.  Maximum disk is not as well
defined a concept in LSB as in HSB galaxies.  The required mass-to-light ratios
are unreasonably high for stellar populations: this just transfers some of the
mass discrepancy from halo to disk.  Even with high disk $\Upsilon_*$, the
halo dominates down to small radii.  

\subsection{Cuspy Halos}

Simulations of structure formation in the CDM cosmogony (e.g., Navarro, Frenk,
\& White 1997; Moore et al.\ 1999) now resolve the structure of individual dark
matter halos.  Though there remains some debate over details, there does now
appear to be widespread agreement that at small radii CDM halos should have a
cuspy distribution ($\al \le -1$).  This is markedly different from the constant
density cores of pseudo-isothermal halos traditionally (and successfully) used
in rotation curve fits.  A cusp has more dark mass at small radii, which reduces
the disk mass which can be simultaneously accommodated.  Cuspy halos and
maximal disks are mutually incompatible. 

\begin{figure}[t]
\psfig{figure=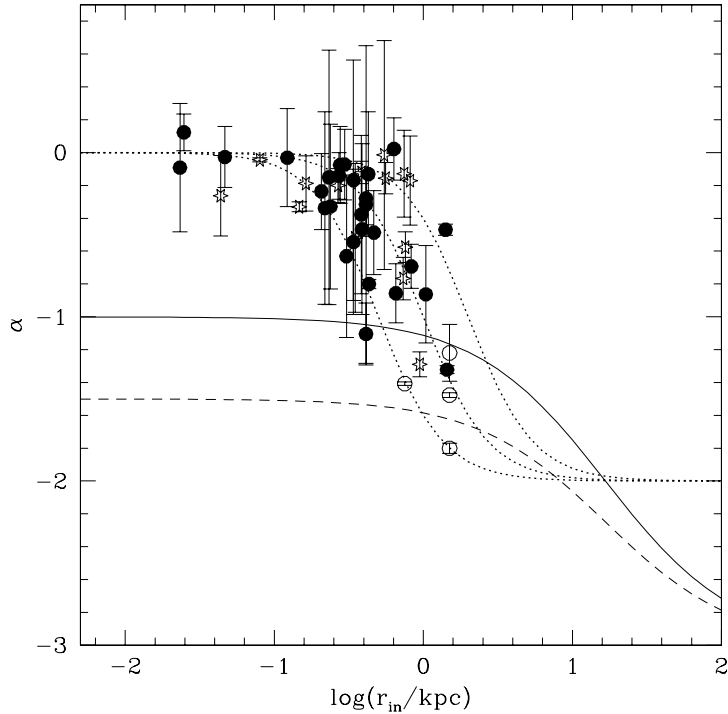,height=4.0in,width=4.0in}
\caption{The cusp slope $\alpha$ 
plotted against the radius of the innermost measured point.
The best resolved data strongly prefer cores (dotted lines) over cusps
(solid and dashed lines).
\label{slope_radius_min}}
\rule{5cm}{0.2mm}\hfill\rule{5cm}{0.2mm}
\end{figure}

The maximum allowance one can make for the cusp is in the limit of zero disk
mass.  This is obviously an unrealistic extreme: stars do have mass.  So does
gas, which can be significant in LSB galaxies (Fig. 2).  However, allowing for a
reasonable amount of baryonic mass limits the room available for a cusp, so as
a conservative limit let us explore the case of zero disk.

Using the most recent high resolution H$\al$ data, we have interrogated the
rotation curves of a large sample of LSB galaxies for the cusp slope they prefer
(de Blok et al.\ 2001).  The median $\alpha = -0.2$: much closer to a constant
density core than to a cusp.  This is in the limit of zero disk mass.  As one
begins to make allowance for the stars, then the amount of rotation attributable
to the dark matter is reduced, further reducing the allowed cusp slope.  

There has been considerable controversy over this issue, with much discussion
of how beam smearing in 21 cm data might hide a cusp (e.g., van den Bosch \&
Swaters 2000).  The new high resolution H$\al$ data address this issue directly
(de Blok et al.\ 2001; see also Blais-Oullette et al.\ 2001; Salucci 2001;
Borriello \& Salucci 2001; C\^ot\'e et al.\ 2000).  Fig.\ 3 shows the cusp slopes
derived from H$\al$ data as a function of physical resolution.  Resolution has the
opposite effect from what is implied by van den Bosch \& Swaters (2000).  It is
only the data which are poorly resolved which are consistent with a cusp. 
Such data are also consistent with a constant density core with a modest
($\sim 1$ kpc) core radius.  On the other hand, those objects which are well
resolved strongly prefer $\alpha = 0$ over $\al \le -1$.  {\it All\/} the data are
consistent with $\al = 0$, while {\it none\/} of the best resolved data tolerate
a significant cusp. 

The inner slope $\al$ shown in Fig.\ 3 has been derived in the limit of zero disk
mass.  Once allowance is made for the stars, the situation for cusps becomes
even worse.  While there may be very good theoretical reasons to expect cuspy
halos, there is no guarantee that reality will be cooperative.  The cusp-core
problem is genuine.

\section{Conclusions}

Rotation curves provide strong constraints on the radial potential in disk
galaxies.  This in turn constrains the mass and distribution of the luminous and
dark components of these galaxies.  Disk masses consistent with those expected
for stellar populations are consistent with the dynamical data, provided halos
have constant density cores rather than cusps.  Cuspy halos require abnormally
low stellar mass-to-light ratios, and are strongly at odds with much of the data
even in the extreme limit $\Upsilon_* \rightarrow 0$.

\section*{Acknowledgments}

I am most grateful to my collaborators, Vera Rubin and Erwin de Blok, for their
long, concerted, and ultimately fruitful efforts.  I thank Albert Bosma and
Jerry Sellwood for their attempts to inject some sanity into the debate
over disk masses and cuspy
halos.  All of us who attended this most entertaining workshop owe a great debt
of gratitude to its organizers, especially Priya Natarajan.
The work of SSM is supported in part by NSF grant AST9901663.

\end{document}